\documentstyle[preprint,aps]{revtex}
\begin{document}
\draft
\widetext

\title{ A model for High Temperature Superconductors using the
Extended Hubbard Hamiltonian}
\author{Evandro V.L. de Mello\cite{email}}
\address{Instituto de F\'{i}sica, Universidade Federal
         Fluminense,\\
         Av. Litor\^anea s/n, Gragoat\'a, 24210-340 Niter\'oi,
         Brazil}
\date{\today}
\maketitle
\begin{abstract}
We derive a method to study the phase diagram for high temperature
superconductors (HTCS). Our starting point is the Hubbard Hamiltonian
with a weak attractive interaction to obtain the formation of bound
pairs. We consider this attractive potential at different positions
for different compounds accordingly to the experimental results of the
coherence length. We then construct a wave function of the BCS type
by a variational method using the Fourier transform of this extended
Hubbard potential and then derive an energy gap equation.
This approach allows us to obtain the critical temperature as function of
the doping concentration which gives very good agreement with the
experimental phase diagrams of  YBaCuO and La(Sr,Ba)CuO compounds.

\end{abstract}

Most of the HTSC are quasi-two-dimensional insulators which become a
metallic conductor and a superconductor below some critical
temperature\cite{Bednorz,Cyrot}. Another property very different from
the usual
superconductors is the very short coherence length of an electron (or hole)
pair, $\xi \approx 10 \AA$. At present, there is no clear consensus about
the origin of the mechanism of attraction but in several proposed
models\cite{Davydov,Micnas,Mott}, the superconductor state is achieved from
a hard core charged boson (formed by  real space pairs)
condensation in analogy with the $^3{He}$ problem.
On the other hand, LDA calculations\cite{Mattheiss}  indicates that
the main features of the $La_{2}CuO_{4}$ band structure can be understood
in terms of a two-dimensional tight-binding model. Another important
point is that the properties of the normal metallic state are different
than those of a common metal described by a free electron gas\cite{Cyrot}.
As it is well known from the study of the Hubbard
Hamiltonian\cite{Anderson}, the on-site Coulomb correlations may explain
the antiferromagnetism at low doping regime, the large magnetic fluctuations
and the semiconductor-like  properties of
the metallic phase. On the other hand, in order to derive the formation of
pairs and their binding energy, a weak attractive interaction $U_1$ may be
added to the Coulomb on-site repulsion $U_0$\cite{Micnas}.
The existence of  bound states suggests that the
normal ground state for many electrons in a tight-binding band may
become unstable in the presence of these interactions $U_0$ and $U_1$.

In this letter, we shall use a BCS type wave function and a variational
method to derive an energy gap or order parameter equation. We shall also
use the intersite attractive potential $U_1$ at different positions than
the usual nearest-neighbor of the extended Hubbard model. These choices of
intersite positions are directly determined from the experimental values
of the coherence length and therefore, they depend on the specific
compound to be studied.
This procedure enable us to obtain the variation of $T_c$ on the hole
concentration which we  compare with the experimental critical
temperatures curves for the YBaCuO and the La(Sr,Ba)CuO compounds.
Thus, let us start considering  the two-dimensional
extended Hubbard model on a square lattice

\begin{equation}
H=-\sum_{\langle ij \rangle , \sigma}-t(c^{\dagger}_{i \sigma} c_{j \sigma}
 + h.c.)+U_0 \sum_{i} n_{i \downarrow}n_{i \uparrow}-U_{1}\sum_{\langle ij
 \rangle}n_{i}n_{j}
\label{Hamiltonian}
\end{equation}
where t is the transfer integral (and the band width is 16t), $U_0$ is the
on-site Coulomb interaction and $U_1$ is an intersite attraction and
$<ij>$ refers to nearest-neighbor pairs.

Let us now study the two-electrons (or two-holes) problem.
In this low-density limit, for s-wave pairs, an exact solution of the
Schr\"odinger equation can be worked out in terms of the lattice Green's
function. The biding energy $\Delta$ for a pair just below the
bottom of the band is given by\cite{Micnas,Mello}

\begin{equation}
tG_{00}(8t+\Delta)=\frac{-U_{1}/2t-U_{0}U_{1}/16t^2-1}{-4U_{1}/t-U_{0}
                   U_{1}/2t^2 +U_{0}/t} .
\label{bound1}
\end{equation}

  To obtain $\Delta$ we make use of an expansion for $G_{00}$ in
terms of an elliptic integral.  In the limit of $\Delta\ll t$, it was
shown that\cite{Bagchi}
\begin{equation}
tG_{00}(8t+\Delta)\approx \frac{1}{2\pi}\left((1.38+0.2\sqrt{
\Delta/t})-(0.25+0.125\sqrt{\Delta/t})\ln(4\Delta/t)\right) .
\end{equation}

These equations allow us to study how $\Delta$ varies with
$U_0/t$ and $U_0/U_1$. In fact, Eq.\ (\ref{bound1}) is valid only
in the low density limit and it
suggests the formation of electron (hole) pairs at the
bottom of the band.
A derivation of Eq.\ (\ref{bound1}),
using real space methods\cite{Mello}, shows that these
electron pairs have center of mass at rest and consequently,
behave like  Cooper pairs with momenta $\vec k$ and $-\vec k$.
After  these preliminary considerations, let us focus on the many-body
problem.  For this purpose, we construct a trial wave function of the BCS
type describing pair of electrons (or holes)

\begin{equation}
\vert \Phi\rangle=\prod_{\vec k}\left(u_{\vec k}+v_{\vec k} a^{\dagger}
_{\vec k \uparrow} a_{-\vec k \downarrow}\right)\vert \Phi_0 \rangle,
\end{equation}
where $\vert \Phi_0\rangle$ is the empty band state and $u^{2}_{
\vec k}+v^{2}_{\vec k}=1$.

Following the variational approach\cite{de Gennes}, we must minimize the
expression

\begin{equation}
\langle \Phi\vert H \vert \Phi\rangle -\mu
\langle \Phi\vert N \vert \Phi\rangle ,
\end{equation}
which yields
\begin{equation}
\langle \Phi\vert H-\mu\vert\Phi\rangle=2\sum_{\vec k}
\xi_{\vec k}v^{2}_{\vec k}+\sum_{\vec k\vec l}V_{\vec k\vec l}
u_{\vec k}v_{\vec k}u_{\vec l}v_{\vec l} ,
\end{equation}
with $\xi_{\vec k}=-4t(cos(k_{x}a)+cos(k_{y}a))-\mu$  and with $V_{
\vec k\vec l }$  being the interaction part of the potential that describes
the transition of a pair from the state $(\vec k,-\vec k)$ to $(\vec l,
-\vec l)$. The minimization procedure follows exactly as the BCS theory
for the free electron gas\cite{de Gennes} and we obtain the same type of
$T=0$ energy gap equation

\begin{equation}
\Delta_{\vec k}=-\sum_{\vec l}V_{\vec k\vec l}\frac{\Delta_{\vec l}}
{2\left(\xi^2_{\vec l}+\Delta^2_{\vec l}\right)^{1/2}} .
\label{gap}
\end{equation}

As we already mentioned, $V_{\vec k\vec l}$ is
the Fourier transform of the potential of Eq.\ (\ref{Hamiltonian}),
which is approximately given by

\begin{equation}
V_{\vec k\vec l}=\lbrack V_0^{1/2}-2V_1^{1/2}\cos(k_x-l_x)a\rbrack
\lbrack V_0^{1/2}-2V_1^{1/2}\cos(k_y-l_y)a \rbrack ,
\end{equation}
or
\begin{equation}
V_{\vec k\vec l}\approx V_{0}-4(V_{0}V_{1})^{1/2}\left(\cos(k_x-l_x)a+
\cos(k_y-l_y)a\right) .
\label{kpot}
\end{equation}

Comparing with the Hamiltonian Eq.\ (\ref{Hamiltonian}), we identify
$U_1=\sqrt{V_0V_1}$ and $U_0=V_0$
and we use that $V_0 > V_1$. Likewise BCS theory, we assume the gap to
have the same functional form of the potential, namely, $\Delta_{\vec k}=
\Delta(0) (cos(k_{x}a)+cos(k_{y}a))/2$. This functional form was also
previously deduced for the RVB particle-particle energy
gap\cite{Baskaran} using self-consistent field methods\cite{de Gennes}.
Although this potential-energy gap relation
is the same used in  the BCS theory, it is worth to stress that the above
potential is very different than
the constant and isotropic mean potential of the original BCS theory
and it has the correlations built on. We take $\mu$  equal to the
hole maximum energy and we will assume that it grows linearly with the
concentration of holes. Although this dependence is an ansatz here, it
has been previously derived by auxiliary-bosons mean field
theory\cite{Baskaran,Kotliar}.

We calculate the probability of finding
a hole pair, that is, the condensation amplitude $F_{\vec k}\equiv
u_{\vec k}v_{\vec k}$. It has a maximum at $k_M$
and drops very rapidly with $\left\vert k\right\vert >k_M$, that is,
the pair formation instability  occurs mostly at the Fermi
surface. This is the same result of the free-electron
BCS theory, despite the fact that
our potential (Eq.\ (\ref{kpot})) acts  on
all the first Brillouin zone. This is mainly due to the small values of
the Fermi surface and the very short coherence length which indicates
that all carriers can be involved in pairing\cite{Micnas}.
According to these considerations and taking $\vec k =0$ in Eq.\
(\ref{gap}), we obtain

\begin{equation}
 1=-\frac{1}{2\pi^2} \int_{0}^{\alpha_\Delta}\int_{0}^{\beta_\Delta}
    \frac{\left(V_0-
    4(V_{1}V_{0})^{1/2}f(\alpha,\beta)\right)f(\alpha,\beta)}{\left(\left(
    4t\left(f(\alpha,\beta)-f(\alpha_M,\beta_M) \right)\right)^2+
    (\Delta f(\alpha,\beta)/2)^2 \right)^{1/2}}\; d\alpha d\beta ,
\label{gap0}
\end{equation}
where $f(\alpha,\beta)\equiv \cos\alpha+\cos\beta$ and
$\alpha=k_{x}a$ and $\beta=k_{y}a$. $\alpha_M$ and $\beta_M$ are the
maximum $T=0$ occupied values (like a Fermi momentum) that
depend on the density of holes or, as more currently used, on the number
of holes per $Cu$ atoms, $x$. Then we can show that
$\alpha_{M}= \beta_{M}\approx arcos(1-x)$. The integrations are
performed up  to $\alpha_\Delta$ and $\beta_\Delta$, which are
chosen at values where the condensation amplitude becomes very
small,  namely, $F_{\vec k} \approx 0.01$. This is
usually attained for $\xi_{\vec k}> 6\Delta(0)$
and larger values of $k$ do not modify the final results.
The two-dimensional integrals are done by
an elementary Simpson's rule algorithm\cite{Press}.
Thus, for given values of $V_1/V_0$ and
$V_0/t$, there are corresponding values of $\Delta (0)$ and clearly
the  relation among them depends on $x$. Furthermore, there may be
only real solutions at a certain range of $x$.

We have derived an expression for the $T=0$ energy gap (Eq.\ (\ref{gap0})).
For $T\not=0$, the
excitations with their respective probability must be taken into
account. The derivation of a self-consistently temperature-dependent
gap equation is analogous to that which leads to Eq.\ (\ref{gap0}).
At this point, we again follow the
BCS approach\cite{de Gennes} and assume that $\Delta(T)$ vanishes
at the critical temperature $T_c$, which yields the following equation
\begin{equation}
1=-\frac{1}{2\pi^2}\int_{0}^{\alpha_\Delta}\int_{0}^{\beta_\Delta}
                       \frac{\left(V_{0}-4\left(V_{1}V_{0}\right)^{1/2}
                       f(\alpha,\beta)\right)f(\alpha,\beta)\tanh
     \left(\frac{2\xi_{\alpha\beta}}{2K_{B}T_{c}}\right)}
 {4t\left(f(\alpha,\beta)-f(\alpha_M,\beta_M)\right)}\; d\alpha d\beta,
\label{gapT}
\end{equation}
where we again integrate up to $\alpha_\Delta$ and $\beta_\Delta$ which
except for $x$ close to one, is almost entirely within the lower Hubbard band.
It is a well known result that the correlation $V_0$ can split the
half-filled conduction band into two, with the upper band being empty
and the lower band being filled\cite{Fulde}. Accordingly the properties of
our equations are also dominated by the lower band. We should also point
out that in the derivation of Eq.(11), we used the entropy form of a
non-interacting fermion system which is an approximation which
is justified in the low energy gap limit, that is, for $\Delta (0)\ll 16t$.

Now we compare Eq.\ (\ref{gap0}) and Eq.\ (\ref{gapT}) in order to
relate $V_1/V_0$, $V_0/t$, $\Delta (0)$ and $T_c$.
The first important thing to notice is that the different coherence lengths
for different compounds suggests that the intersite attractive pair
potential
should be placed at a different site than the original nearest neighbor
site of the extended Hubbard Hamiltonian. In other words, the
minimum of the real attractive potential depends on the type of HTCS.
This can be easily included in our equations by a
change in the lattice parameter only in the potential (Eq.\ (\ref{kpot})
and $\Delta_{\vec k}$ expressions. Thus,
for the lanthanum compounds, since the lattice parameter is approximately
$4\AA$ and  $\xi \approx 35\AA$ \cite{Cyrot} , we expect that
the minimum of the attractive pair potential  occurs at a distance of the
order of $8-9$ lattice parameters.
Similarly, with a coherent length $\xi \approx 15\AA$ \cite{Cyrot}, we expect
the attractive pair potential minimum to be  at  $3-4$ lattice
parameters for the yttrium
compounds. Thus, in order to obtain values pertinent to the experimental
results for the $La_{2-x}Sr_{x}CuO_{4}$ compounds, we use
the attractive potential  at the
$6^{th}$-neighbor position and with $V_0/V_1=8^2$. We find a real
solution only for values of $x\in [0.04,0.35]$ and with a maximum value for
$T_c$ at $x=0.16$. Thus, the position of the maximum is fixed and the
absolute values of $T_c$ depend
only on $\Delta(0)$. Therefore, to compare with the measurements, we use
$16t=2 ev$\cite{Mattheiss} and we choose
$\Delta(0)=80K$ in order to obtain  $T_c=35K$ at $x=0.16$.
In Fig.1, we plotted our values for $T_c$ as function of $x$ and we
see that they provide an excellent fit for the
experimental data points which were taken from Refs.\cite{Micnas,Mott}.
Furthermore, we obtain  values for
$\Delta(0)\over{K_BT_c}$ greater than 2.3 (around $x=0.16$) which agrees with
earlier measurements\cite{Muller} which yielded values larger than
2.5. These results are considerably larger than the BCS value of 1.75.

For $YBa_{2}Cu_{3}O_{6+x}$ we place the attractive potential at the
next neighbor position. With $V_0/V_1=5^2$, we find  that there are
solutions only for $x>0.4$ and the maximum $T_c$ occur at $x=0.96$,
which is close to the measured maximum $T_c$ at $x=0.93$. Now we choose
$\Delta(0)=180K$ which gives $T_c=95K$ at $x=0.96$. Our results are
plotted in Fig.2 with also some experimental data points taken from
Ref.\cite{Cyrot}. We notice that the agreement in this case is only
qualitative and fail to reproduce the change in concavity around
$x=0.7$. A possibility is that this structure is due to inter-plane
effects in the compounds richer in oxygen and, therefore, it should not
be reproduced by our two-dimensional treatment.
A difficulty with our calculations is the variation of the
values of $V_0/t$ (and $V_1/t$) with $x$. For the lanthanum
compounds, as a result of Eqs.(10) and (11), we obtain $V_0/t=11$
for $x=0.16$ and at the onset of superconductivity, we find
$V_{0}/t\approx 11$ (at $x=0.05$). Although they are in the same
order of magnitude of the expected values of
$V_0/t=16$\cite{Fulde}, such variation of the coupling with the
concentration is artificial. It is probably due to the linear
approximation used for the chemical potential.

In summary, we demonstrated that a variational procedure with a
BCS type wave function in connection with a two-dimensional Hubbard
Hamiltonian and with an attractive interaction is  suitable to
determine the superconducting state for layered HTSC.
This approach allows us to  derive the energy gap equation in $\vec
k$-space, which is used to relate $T_c$ with $x$. The results yield an
excellent fitting for the curves of the critical temperature as
function of doping when we take an effective position for the attractive
potential at a distance consistent with the measured values for the
coherence length.
The different positions for the attractive potential are probably due to the
coupling to a bosonic field (phonons, excitons, plasmons, etc....). Such
field
is polarized along the electronic motion and induces an effective short
range attraction (renormalized from its bare value) which varies for
different compounds. Other experimental values, like the ratio
$\Delta/T_c$, are well reproduced by our method. Thus, we conclude that the
method derived here can be  successfully applied to the two most studied
compounds, even
though, they have  transitions at very different ranges of $T$ and $x$.

The author acknowledges fruitful discussions with prof. Mucio Continentino
and prof. Mauro Doria. He also thanks CNPq and FINEP for partial financial
support.

\begin{figure}
\caption{ The critical temperature for  $La_{2-x}Sr_xCuO_4$ as function  of
the hole concentration $x$. The solid line is plotted from our calculations
and the black dots are experimental results taken from
Refs.4 and 5 .}
\label{fig1}
\end{figure}

\begin{figure}
\caption{ The critical temperature for   $YBa_2Cu_3O_{6+x}$ as function of
the hole concentration $x$. The solid line is plotted from our calculations
and the black dots are experimental results taken from a plot in
Ref.2 . }

\label{fig2}
\end{figure}
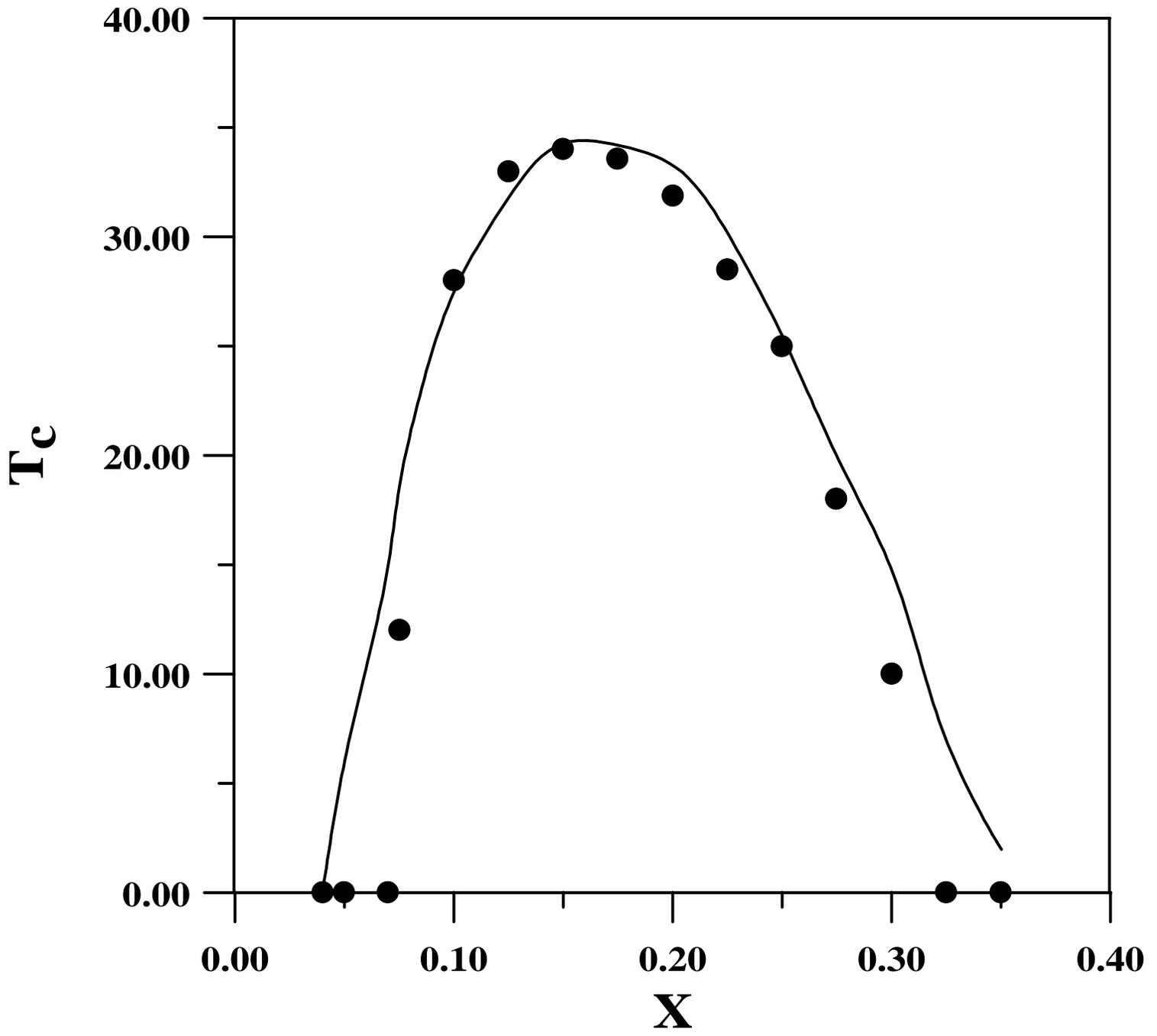
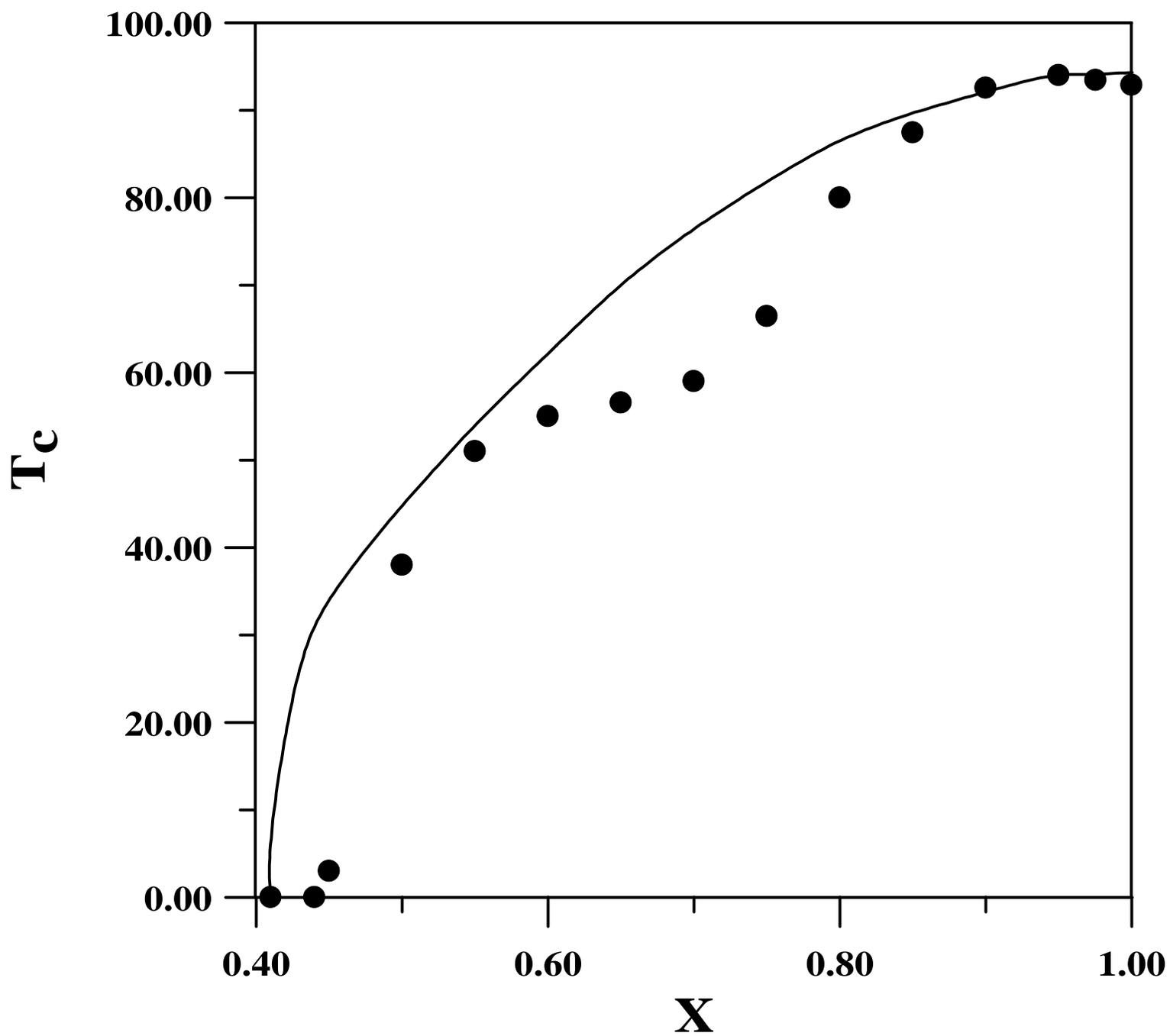
\end{document}